\documentclass{sig-alternate}

\usepackage{nth}
\usepackage{ulem}
\usepackage{cleveref}
\usepackage{enumerate}
\usepackage{xcolor}
\usepackage{etoolbox}

\usepackage[shortlabels]{enumitem}
\setlist[enumerate,1]{leftmargin=0.4cm}

\colorlet{punct}{red!60!black}
\definecolor{background}{HTML}{EEEEEE}
\definecolor{delim}{RGB}{20,105,176}
\colorlet{numb}{magenta!60!black}

\usepackage{listings}
\lstdefinelanguage{json}{
    basicstyle=\scriptsize\ttfamily,
    numberstyle=\scriptsize,
    stepnumber=1,
    numbersep=8pt,
    showstringspaces=false,
    breaklines=true,
    frame=lines,
    backgroundcolor=\color{background},
    literate=
     *{0}{{{\color{numb}0}}}{1}
      {1}{{{\color{numb}1}}}{1}
      {2}{{{\color{numb}2}}}{1}
      {3}{{{\color{numb}3}}}{1}
      {4}{{{\color{numb}4}}}{1}
      {5}{{{\color{numb}5}}}{1}
      {6}{{{\color{numb}6}}}{1}
      {7}{{{\color{numb}7}}}{1}
      {8}{{{\color{numb}8}}}{1}
      {9}{{{\color{numb}9}}}{1}
      {:}{{{\color{punct}{:}}}}{1}
      {,}{{{\color{punct}{,}}}}{1}
      {\{}{{{\color{delim}{\{}}}}{1}
      {\}}{{{\color{delim}{\}}}}}{1}
      {[}{{{\color{delim}{[}}}}{1}
      {]}{{{\color{delim}{]}}}}{1},
}

\newcommand{\framework}{\textsc{Seastar}}

\usepackage{eso-pic}

\usepackage{cleveref}
\crefname{paragraph}{\S}{\S\S} 

\newcommand{\suppress}[1]{}  

\AddToShipoutPicture{%
           \AtPageCenter{%
        \put(120,-500){\rotatebox[origin=c]{45}{\parbox[t]{10cm}{\color[gray]{0.80}%
        \scalebox{14}{DRAFT}\par
    }}}}%
}

\makeatletter
\def\@copyrightspace{\relax}
\makeatother

\begin{document}

\title{Towards a Comprehensive Framework for Telemetry Data
in HPC Environments}

\numberofauthors{3}

\author{
\alignauthor Ole Weidner\\
    \affaddr{School of Informatics}\\
    \affaddr{University of Edinburgh, UK}\\
    \email{ole.weidner@ed.ac.uk}
\alignauthor Malcolm Atkinson \\
    \affaddr{School of Informatics}\\
    \affaddr{University of Edinburgh, UK}\\
    \email{malcolm.atkinson@ed.ac.uk}
\alignauthor Adam Barker\\
    \affaddr{School of Computer Science}\\
    \affaddr{University of St Andrews, UK}\\
    \email{adam.barker@st-andrews.ac.uk}
}

\date{01 February 2016}

\maketitle

\begin{abstract}

  A large number of \textit{\nth{2} generation} high-performance computing
  applications and services rely on adaptive and dynamic architectures and
  execution strategies to run efficiently, resiliently, and at scale on today's
  HPC infrastructures. They require information about applications and their
  environment to steer and optimize execution. We define this information as
  \textit{telemetry data}.

  Current HPC platforms do not provide the infrastructure, interfaces and
  conceptual models to collect, store, analyze, and access such data. Today,
  applications depend on application and platform specific techniques for
  collecting telemetry data; introducing significant development overheads that
  inhibit portability and mobility. The development and adoption of adaptive,
  context-aware strategies is thereby impaired. To facilitate \textit{\nth{2}
  generation} applications, more efficient application development, and swift
  adoption of adaptive applications in production, a comprehensive framework for
  telemetry data management must be provided by future HPC systems and services.

  We introduce \framework{}, a conceptual model and a software framework to
  collect, store, analyze, and exploit streams of telemetry data generated by
  HPC systems and their applications. We show how \framework{} can be integrated
  with HPC platform architectures and how it enables common application
  execution strategies.

\end{abstract}

%
%

%
%
\printccsdesc



\section{Introduction}
\label{sec:intro}

With computational methods, tools and workflows becoming ubiquitous in more and
more disciplines, the software applications and user communities on HPC
platforms are rapidly growing diverse. Many of the \nth{2} generation HPC
applications~\cite{weidner2016rethinking} have moved beyond tightly-coupled,
compute-centric methods and algorithms and embrace more heterogeneous,
multi-component workflows, which involve adaptive, dynamic, computation and
data-centric methodologies. While diverging from the traditional HPC application
profiles, many of these applications still rely on the large number of tightly
coupled cores, cutting-edge hardware and advanced interconnect topologies
provided by HPC clusters. Examples of \nth{2} generation applications are
user-level scheduling frameworks like pilot jobs, and applications with dynamic,
or hard-to-predict runtime trajectories like Kalman Filter and Adaptive Mesh
Refinement (AMR) applications.

The more traditional HPC applications and frameworks like MPI have also started
to explore adaptive techniques to scale up on platforms that are continuously
growing in capacity. For these applications, running at extreme scales bears a
twofold risk: a statistically increased risk of hardware and software failure,
and increasing costs in case of application failure. Implementing adaptivity and
resilience can alleviate these risks. For example, an application that
understands its performance profile with a given configuration might decide to
terminate early or adjust when it detects inefficient execution, e.g., due to
excessive swapping or slow I/O.

Most of these dynamic and adaptive techniques require the applications to have a
model about themselves (self aware) and their environment (context aware). With
such a model, applications can implement mechanisms like feedback loops to
validate their execution parameters and trajectory, and to react and adjust
according to their objectives.

Telemetry data is the continuous streams of run-time information that is
generated by HPC systems, and the services and applications running on them. It
includes operating system metrics at the process, and thread level, metrics
describing the state of I/O resources, network interconnects, and storage
facilities, as well as metrics describing the state of job schedulers and other
HPC services. In short, telemetry data integrates all the information that is
generated \textit{about} platforms and applications. It is distinct from the
data that is generated \textit{by} the applications, which we refer to as
application data.

Existing approaches to context awareness and management and provisioning of
telemetry data are scattered throughout the application and infrastructure
landscape. None are comprehensive across platforms, environments and
applications. This causes significant development overheads, with duplication of
localized solutions that reduce portability and mobility. It impedes the
development and adoption of adaptive, context aware strategies and
architectures. From our perspective, a comprehensive and unifying framework for
telemetry data management must be provided by future HPC platforms as a  system
service to facilitate a more efficient application development lifecycle, and a
swift adoption of adaptive application research into production.

\subsection{Approach and Contributions}
We propose a solution to the provisioning and integration of telemetry data on
HPC platforms. This is important and timely because an increasing number of HPC
applications rely on it to implement context aware, dynamic and adaptive
execution strategies. We are not aware of any other solution emerging. This
paper introduces \framework{}, a model, API, and implementation blueprint that
facilitates the collection, management and use of telemetry data on HPC
platforms, and simplifies the development of context aware HPC applications.
This paper makes conceptual and practical contributions to HPC platform and
application design:

\begin{enumerate} [itemsep=0mm]


  \item{It develops a graph-based model called \framework{} that allows to
  capture telemetry data within a dynamic graph that represents the
  continuously changing application and platform structure of an HPC cluster.
  }

  \item{It defines a programming interface (API) for applications and system services
  to query and analyze platform and application structure and telemetry data as
  a core concept to simplify the development of adaptive applications (\cref{sec:api}).
  }

  \item{It describes an architecture blueprint for a framework that
  implements \framework{} on an existing HPC cluster (\cref{sec:implementation}).
  }

\end{enumerate}

%
This paper is structured as follows: \Cref{sec:background} discusses concepts
related to telemetry data. \Cref{sec:usecases} presents application
use-cases and challenges. \Cref{sec:model} introduces \framework{}, a
graph-based model that captures and organises telemetry data. \Cref{sec:api}
describes the API for applications and platform services to interact with
telemetry data. \Cref{sec:implementation} lays out a blueprint for an
implementation of \framework{}. Section~\ref{sec:conclusion} presents plans to
evaluate \framework{} and discusses future research into telemetry data management
at scale and in distributed contexts.

%
%
\section{Background}
\label{sec:background}

In~\cite{weidner2016rethinking} we have argued that bringing together
application developers with HPC-resource providers on both technical and
cultural levels is a big challenge with substantial \textit{potential} benefits.
The prevailing separation between the two communities is the main cause for the
lack of interfaces and information flow across the application-platform divide.
Similar observations can be found in~\cite{fialho2014framework} where Fialho
\textit{et al.} point out a lack of a common frameworks for telemetry data as many HPC
performance optimization tools implement some or several aspects of the full
performance optimization task but almost none are comprehensive across
architectures, environments, applications, and workloads. Similarly,
{\'A}brah{\'a}m \textit{et al.}~\cite{abraham2015preparing} propose methodologies to
efficiently collect run-time information as a preparation for autonomic exascale
applications.

\subsection{Application Areas}
\label{sec:usecases}

Use-cases for telemetry data are manifold and an exhaustive survey would not be
feasible in this context. Here we lay out six high-level application areas
for telemetry data in HPC along with brief examples to illustrate the broad
landscape of telemetry data usage.\smallskip{}

\noindent{}\textbf{Application Development Lifecycle} is an iterative
process from concept to production. It requires profiling, collecting
information  about performance data, networking, and I/O patterns so that the
application developer can decide between alternatives or fine-tune for a specific
architecture. Profiling data is collected by instrumenting either the program
source code, its binary executable, or its run-time environment. Especially
during the development of large-scale parallel code, profiling tools like e.g.,
Vampir/NG~\cite{brunst2003distributed}, PAPI~\cite{browne2000portable}, and
TAU(g)~\cite{huck2006taug} play a critical role in the optimization
process. While all these tools collect large amounts of telemetry data,
the data is not accessible outside these frameworks or programmatically
during the runtime of an application.\smallskip{}

\noindent{}\textbf{Adaptive Applications} have many application areas.
Some of the more prominent examples are Adaptive Mesh Refinement (AMR) and
Kalman-Filters which exhibit hard-to-predict execution trajectories
and heterogeneous computational loads. When these are ignored, the performance
of these applications can suffer significantly. Adaptivity is also needed to
handle external factors, e.g., Eisenhauer \textit{et al.}~\cite{eisenhauer2009event} have
shown how one application's massive I/O operations perturb the performance of
other applications on the same system. Telemetry data is critical to implement
adaptivity. \smallskip{}

\noindent{}\textbf{Adaptive Runtime Systems} provide low-level load balancing
and scaling capabilities for parallel and distributed applications. Adaptive
MPI~\cite{huang2004adaptive} for example is an alternative run-time for MPI
applications. Charm++~\cite{kale1993charm++} and
Parallax/HPX~\cite{kaiser2009parallex} provide their own programming models and
APIs. All frameworks collect telemetry data via operating system interfaces and
evaluate them via a performance model to make (re-)scheduling decisions.
However, the model and associated data is generally not easily accessible
externally.\smallskip{}

\noindent{}\textbf{Computational Steering} allows applications to be dynamically
configured (steered) at run-time; as opposed to adaptive run-time systems where
adaptivity is transparently provided by the underlying framework. Here the
feedback loop is moved into the application space, which also requires context
data available in application space. Hence steering frameworks often have a
monitoring component, e.g., FALCON~\cite{gu1995falcon}, an on-line monitoring
and steering framework for large-scale parallel applications, and
\cite{eisenhauer1998object} an object-based infrastructure for program
monitoring and steering.\smallskip{}

\noindent{}\textbf{Resource Aware Scheduling} allows the (re-)scheduling of HPC
workloads based on the observed resource utilization. I/O aware
scheduling~\cite{zhou2015aware} for example, can control the status of jobs on
the fly during execution based on run-time monitoring of system state and I/O
activities. Another example is the \textit{COBALT}
scheduler~\cite{tang2010analyzing}. In comparison, most existing HPC job
schedulers employ a static, a priori performance model. Fluctuations in the
performance metrics of a resource, e.g., disk or network I/O hotspots are not
monitored or acted upon. While this works well with static and homogeneous
workloads, it fails with the increasing presence of \nth{2} generation
applications.\smallskip{}

\noindent{}\textbf{Application-Level Scheduling} is a tactic to circumvent the
static constraints and granularity of HPC job schedulers. A commonly used method
is to employ \textit{pilot jobs} or ``placeholder jobs'' submitted as a single
job to the job scheduler. Once they are active they accept user jobs that are
then executed within the placeholder job. Examples of application-level
scheduling frameworks are HTCondor~\cite{thain2005distributed} and RADICAL
Pilot~\cite{merzky2015radical}. Most application-level scheduling systems
collect telemetry data via operating-system interfaces to determine how to
schedule their computational workload most efficiently and to detect errors.

\subsection{Context Awareness}

The term context awareness is often used in close proximity with monitoring
and telemetry data. If we look again at the application areas in~\cref{sec:usecases},
all of them require some understanding of the HPC platform context, whether
it is information about other applications running, the execution environment
or the state of the platform and its components.
The development of context-aware applications gained significance with
the emergence of grid computing in the early 1990s when
application developers and scientists had access to a growing
distributed  ecosystem of computational resources and federated HPC systems.
While grids strove to unify access, job submission, and file transfer across
systems, they did not provide abstractions for the different execution
environments. Heterogeneity across hardware architectures, cluster and network
configurations, parallel run-time environments and software stacks made it very
difficult to develop applications that ran well at multiple sites. Consequently,
methods and mechanisms were implemented to detect properties of the system an
application was running on and set application parameters accordingly.
Context awareness is not used consistently in the literature. We offer our
own definition to avoid ambiguity.
Our definition uses the fundamental building
block of the executable representation of an application: the operating system (OS) \textit{process}.
An HPC application consists of many, potentially communicating processes. Their
composition and properties change throughout the application's life-/run-time.
Together with the related terms, \textit{self awareness} and \textit{location awareness},
our working definition of context awareness is as follows:\smallskip{}

\noindent{}\textbf{Self Awareness:}
An application is \textbf{logically} self aware if it collects information about
its application-level structure, properties, and data with the aim to use these
information to  control and optimize its internal processing workflows, algorithms,
etc. An application is \textbf{physically} self aware if it collects information
about of its OS process structure and properties.\smallskip{}

\noindent{}\textbf{Location Awareness:}
An application is location aware if it has a model to \textit{understand} of
the spatial mapping of its processes within the HPC platform.\smallskip{}

\noindent{}\textbf{Context Awareness:}
An application is context aware if it is location aware and has an
\textit{understanding} of the properties of the executing platform
and can correlate these with its own properties.

\subsection{HPC System Monitoring}

System monitoring is at the heart of most HPC systems. It allows system
administrators to have a high-level overview of the entire system and to
identify potential issues and bottlenecks. A problem with system monitoring in
HPC is that it is often considered an administrative tool and not exposed to
users and applications. One of the most widely used monitoring systems is
Ganglia~\cite{massie2004ganglia}, a client-server system that extracts telemetry
data from node operating systems and hypervisors. While data in Ganglia is
internally represented in XML, it is normally available only as pre-rendered
graphs rather than programmatically. Ganglia does not have the notion of an
application, which makes it difficult to correlate application behavior with
observed metrics.

New monitoring systems and tools have evolved in the context of cloud
computing. Naturally, cloud resources are treated as ephemeral and their
performance can fluctuate due to both, internal as well as external factors.
Hence, system monitoring has emerged as an important pillar for cloud
applications and infrastructure. Important tools in this area are Amazon AWS
CloudWatch~\cite{cloudwatch2006online} and
Prometheus~\cite{prometheus2016online}.
As opposed to the monitoring systems found on HPC platforms, these systems
provide extensive APIs that can easily be consumed by applications and other
system services. However, neither of the two system captures the structure
of the underlying platform.

\section{Challenges and Motivation}

As diverse as the application areas for telemetry data, as diverse are the
approaches for its management. From this diversity arises a number of
challenges towards a comprehensive, unified framework for telemetry data
management in HPC environments. In this section we list the ones we consider
most important along with a specific use-case that has motivated our
research in this area.

\subsection{Challenges}

From the application areas and use-cases we have identified a set of challenges
and shortcomings related to operation telemetry data management:\smallskip{}

\noindent{}\textbf{Data Access:} Applications access operating
system facilities, such as the Linux \texttt{/proc} file-system, and
sometimes higher-level interfaces to extract telemetry data. None of
these interfaces are entirely consistent across platforms and operating systems
which introduces portability issues. In addition, many of the interfaces are
relatively low-level which can pose additional hurdles in the development
process. \smallskip{}

\noindent{}\textbf{Historical Data:} Existing operating system interfaces
only provide \textit{ad hoc} data. If HPC applications require historical
telemetry data, e.g., to analyze previous or similar runs, they need to collect
and store this data themselves. \smallskip{}


\noindent{}\textbf{Data Contextualization:} Just looking at telemetry data in
isolation is not sufficient to understand the behavior of an application or
system. The data needs to be interpreted in its context. Application performance
data like network and filesystem I/O, can only be interpreted if we have an
understanding of the  properties of the underlying hardware and software stack,
as well as an understanding of the other actors sharing the same resources.
Similarly, the more information that is made available about the running
applications the better the interpretation of the behavior of hardware and
system services.\smallskip{}

\noindent{}\textbf{Data Correlation:} It is often not feasible to collect all
 telemetry data that is necessary to contextualize a set of metrics in
the same context. Some metrics can only be collected in an application context,
others might be only accessible through a system service. In order to correlate
data that is generated by different, independent entities, a common spatial and
temporal reference system is required. In order to correlate for example the I/O
throughput of a specific operating-system thread with the status and load of
distributed file-system partition, information about the locality of the thread
is required.\smallskip{}


\noindent{}\textbf{Data Analysis:} The volume of telemetry data can become
quickly very large at scale. This makes it difficult to analyze, especially on
the application-side. For example, trying to find suspicious I/O patterns in an
application running across 10,000 processes is not a trivial endeavor. None of
the analyzed systems provide or can make use of analytics facilities that would
allow them to derive high-level signals from a high-volume stream of complex
input data.\smallskip{}

\subsection{Motivating Use-Case}

We use the RADICAL-Pilot~\cite{merzky2015radical} pilot job system to develop
bioinformatics workflows. Many of these workflows spawn large numbers of
short-running processes that can exhibit highly irregular I/O and computation
patterns. Confined to the static resources allocated by HPC schedulers, we use
pilot jobs to (re-)schedule workflow tasks based on their actual behavior and
communication requirements. Furthermore, we want to circumvent system issues
like filesystem I/O and network bottlenecks, which seem to occur in a
surprisingly consistent frequency due to other applications running in the same
vicinity. Lastly, we want to capture and catalog the execution trajectories and
properties of all our workflows to be able to make predication about the
behavior of similar workloads. While RADICAL-Pilot provides effective mechanisms
to run many jobs within a single HPC queueing system job, it does not provide
any convenient mechanisms to collect the telemetry data required. We explored
multiple different ways to collect this data as part of the application logic.
The overhead and inefficiency encountered in the process, especially at larger
scales, required us to take a step back and think about what would be required
to support applications like ours. \framework{} is the direct outcome of this.

\section{Seastar Model}
\label{sec:model}

To provide a generic model to capture telemetry data on an HPC
platform, we define a set of requirements from which we then derive the
graph-based \framework{} model. The overarching goal is not to introduce yet
another platform- or application-specific framework orthogonal to already
existing approaches. Instead, we strive to develop  a generic framework that is
(a) agnostic, i.e., applicable to a broad set of HPC applications and platforms,
and can (b) incorporate existing data sources and put them into a common
context. We define the following requirements:

\begin{enumerate} [itemsep=0mm]

  \item{The model must capture the physical representation (the anatomy) of an
  application, i.e., its processes, threads, and the interdependencies between
  them.}

  \item{The model must capture the layout (anatomy) of the platform, i.e., its
  hardware components, and the interdependencies between them.}

  \item{The model must capture the mapping between the application and the
  platform anatomies, i.e., the physical application representation
  \textit{within} its platform context.}

  \item{Different actors are interested in different aspects of the system. The
  model must support structure and data at an arbitrary level of
  detail.}

  \item{Depending on the use-case, current (live) and / or previous (historic)
   data might be required. The model must capture both.}

\end{enumerate}

\noindent HPC applications span a wide area of categories, ranging from tightly-coupled
parallel applications to distributed workflows and service-oriented
architectures. Each class of application has its own internal logical
representation, concepts and building blocks. The only commonality that exists
across all applications is that once they run, they have the same physical
representation. The physical representation of applications and platforms, i.e.,
their anatomies serve as the starting point for our model definition. For the
application anatomy, we assume a time-variant network of communicating
processes. Each process and communication link can be split up into hierarchical
networks of sub-components. We make an analogous assumption for the platform
anatomy. We make the following assumptions for the \framework{} model:

\begin{enumerate} [itemsep=0mm]

  \item{The physical anatomy of an application can be described as
  nested, hierarchical networks of connected entities.}

  \item{The physical anatomy of an application can change during its lifetime.}

  \item{The anatomy of an HPC platform can also described as
  nested, hierarchical networks of connected entities.}

  \item{The anatomy of an HPC platform can change during its lifetime.}

  \item{The context of an application is defined as its locality within
  an HPC platform, i.e., the mapping of an application anatomy to a platform
  anatomy.}

  \item{The context of an application can change during its lifetime.}

\end{enumerate}

\noindent Based on these assumptions, we define a graph-based representation of
applications and platforms. It consists of multi-layer, directed \textit{anatomy
graphs} that represent applications and platforms. Vertices and edges of anatomy
graphs can hold an arbitrary number of time-series attributes that represent
observed  telemetry data. A mapping of the application anatomy graphs to a
platform graph, called the \textit{context graph}, represents the time-variant
localities of applications within a platform (\cref{fig:anatomy_graphs}).

\subsection{Anatomy Graphs}

Anatomy graphs capture the changing anatomies of applications ($AAG$) and the
HPC platform ($PAG$). They are the foundation for the context graph, which
captures the mapping between ($AAG$s) and ($PAG$). Anatomy graphs are nested
directed graphs which represent application components (vertices) and the
connection between them (edges). Each vertex and edge can have an arbitrary
number of attributes that represent a time series of  data that can be
associated with it. Vertices can have pointers to a nested graph that represents
its parent component at a finer level of granularity. Nesting is strictly
hierarchical: edges can only connect vertexes within the same (sub-) graph.
Connecting the vertices of subgraphs with different parent edges is not allowed,
even if the subgraphs are at the same hierarch depth. Anatomy graphs can be
conveniently written as typed and attributed
\textit{E-Graphs}~\cite{ehrig2004fundamental}: 

\begin{equation*}
AG = (V_{g}, V_{d}, E_{g}, E_{na}, E_{ea}, (source_i, target_i)_{i=1,2,3}),
\end{equation*}

\noindent with graph nodes $V_{g}$ and data nodes $V_{d}$, graph edges $E_{g}$, node
attribute edges $E_{na}$, and edge attribute edges $E_{ea}$, and source and
target functions:

\begin{equation*}
\begin{split}
– source_1 :E_{g} \rightarrow V_{g},\ source_2 :E_{na} \rightarrow V_{g},\ source_3 :E_{ea} \rightarrow E_{g} \\
– target_1 :E_{g} \rightarrow V_{g},\ target2 :E_{na} \rightarrow V_{d},\ target_3 :E_{ea} \rightarrow V_{d}
\end{split}
\end{equation*}

\noindent We amend the \textit{E-Graphs} definition in~\cite{ehrig2004fundamental} so that
data nodes ($V_{d}$) can be a pointer to another (nested) anatomy graph
$AG_{n}$. To capture the potential changes in application and platform anatomy
over time, $AG$ is time-dependent:

\begin{equation*}
  \begin{split}
    AAG(t) = \\
    (V_{g}(t), V_{d}(t), E_{g}(t), E_{na}(t), E_{ea}(t), (source_i, target_i)_{i=1,2,3})
  \end{split}
\end{equation*}

%
%
%

\noindent Figure~\ref{fig:anatomy_graphs} shows and example of application and platform
anatomy graphs. Anatomy graphs allow us to capture a complete picture of the
changing structures of applications and HPC platform. By changing the time
parameter $t$ for an $AAG(t)$, we can ``navigate'' back and forth in the
evolution of an application from beginning (startup) to end (termination). The
ability to track the anatomy of an evolving application is very important for
the post mortem and ad-hoc analysis and optimization of dynamic applications and
task scheduling frameworks.


\subsection{Context Graph}

\begin{figure}[t]
  \includegraphics[width=\columnwidth]{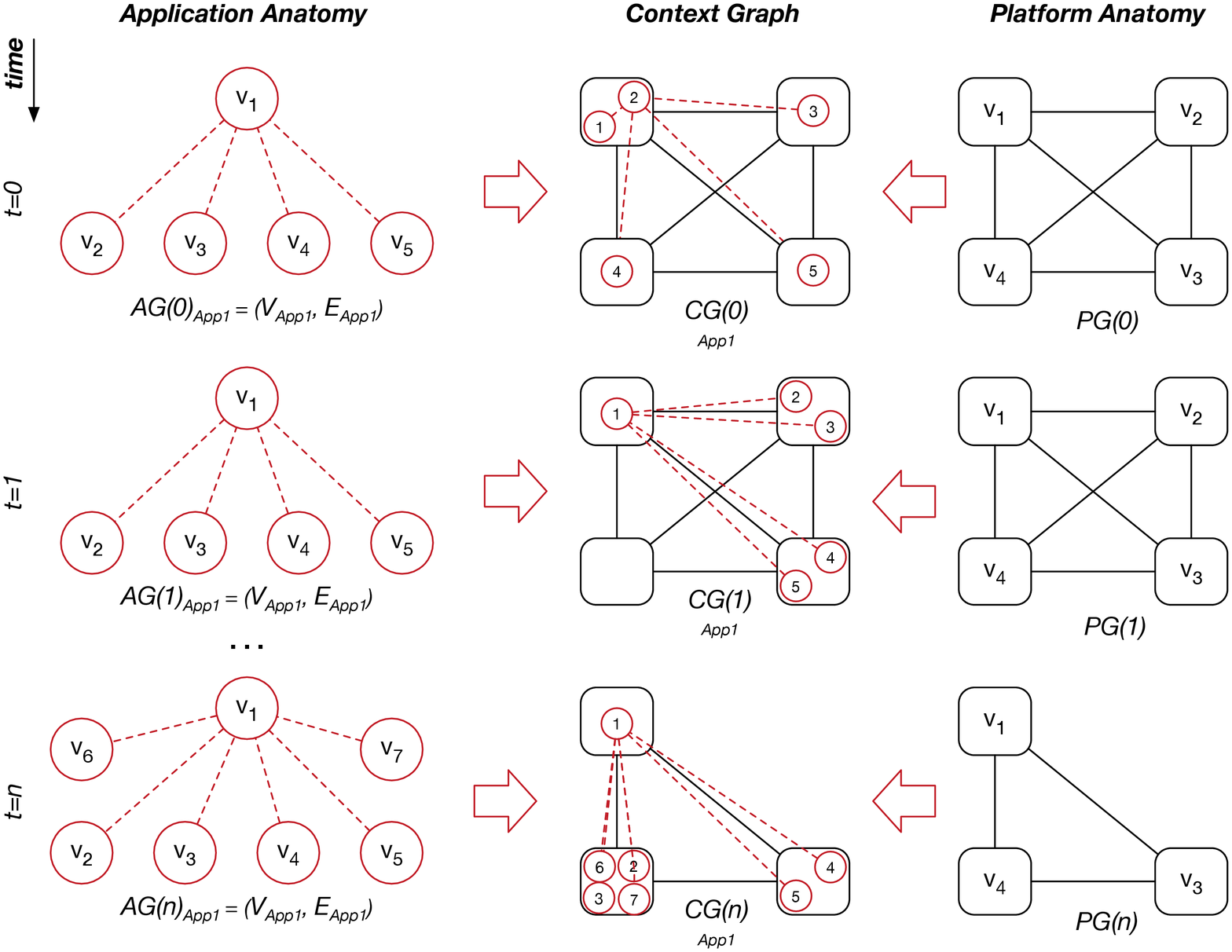}
  \caption{A context graph maps the spatial-temporal application anatomy graphs
  to the spatial-temporal platform graphs. Each instance of a context graph
  captures the structure and properties of applications and platforms at
  a given instant.}
  \vspace{-1.0em}
  \label{fig:anatomy_graphs}
\end{figure}

Context graphs (\cref{fig:anatomy_graphs} r.) capture the time-varying
relationship between a platform anatomy graph and application graphs.

The locality of all applications $AG_{App_1..A_n}(t)$ within the platform
$AG_{P}$ is captured through a fixed mapping function ($\bullet$). We define the
resulting graph as the \textit{global context graph} ($CG_{Global}$) (see Figure
\ref{fig:anatomy_graphs} c.):

$$ CG_{Global}(App, P, t) = AG_{P} \bullet AG_{App_1..App_n}(t) $$

\noindent Additionally, we define application-specific context graphs ($CG_{App_n}$) as
sub-graphs of $CG_{Global}$:

$$ CG_{App_1}(App_1, P, t) = AG_{P} \bullet AG_{App_1}(t) $$

\noindent This spatio-temporal representation creates a set of graph structures in which
the individual components and their mappings can be attributed with context
information.

We can think of the vertices of an application graph ($V_{App}$) as the
operating system processes comprising an application and of the platform graph
vertices ($V_{P}$) as the physical or virtual nodes of an HPC cluster. The edges
can then represent communication between processes ($E_{A}$) and network links
between nodes ($E_{App}$) respectively.

\subsection{Time-Series Data}

Telemetry data, e.g., operating system metrics, is captured as
time-series data and attached to the node and edge attributes of the graphs.
Currently, the \framework{} model does not make assumptions about this data.
Timestamps are set by the entity collecting the data. On an implenentation
level, this assumes that all HPC platform components (nodes) use the same,
synchronized timebase.

%
%

%
%
\section{Seastar API}
\label{sec:api}

\framework{} provides the structure to capture telemetry
data in a graph-based model. The \framework{} API allows applications, platform
services and human actors to explore and interact with this model. The API uses
a RESTful representation and the JSON format to describe return objects. The
return object structure is that of an attributed graph or edge node. From each
node, the hierarchical graph can be traversed via \texttt{parent\_nodes},
\texttt{child\_node}, and \texttt{sibling\_nodes}. A \texttt{timestamp} field
positions the object in temporal space. Attributes describing edge connections
between siblings, e.g., the communication between two MPI processes follow the
the same pattern.

\begin{lstlisting}[language=json, caption={JSON resource object structure}]
 { timestamp: 1491830507,
   parent_node: {
     job: <id>
   },
   child_nodes: {
     threads: []
   },
   sibling_nodes: {
     processes: []
   },
   attributes: {
     m1: [], m2: [], ...
   }
 }
\end{lstlisting}

\noindent The current iteration of the API defines only a subset of possible
resource types but it can easily be extended to additional types and
hierarchies. For application graphs, \texttt{job}, \texttt{process}, and
\texttt{thread} are defined. For the platform graph \texttt{node},
\texttt{processor}, and \texttt{core} are defined.

\subsection{Model Queries}

The API uses \textit{GraphQL}~\cite{graphql2017online} as the query language
to the context graph hierarchies. GraphQL allows the caller to extract
complex structures from the model in a single API call.

\begin{lstlisting}[language=JSON, caption={Get memory consumption of all sibling processes of a job via a GraphQL query.}]
  {
   process(id: 1) {
     siblings {
       processes {
         memory_uses
       }
     }
   }
  }
\end{lstlisting}


\subsection{Context Awareness}

Context awareness requires self awareness and location awareness. Self awareness
can be established via the special \texttt{self} path element. In the current
iteration of the API it can be called on a job, process, or thread resource and
returns the appropriate object for the application from which it was called.

\begin{lstlisting}[language=bash, caption={Self awareness via \texttt{self}}]
 GET /job/self
 GET /process/self
 GET /thread/self
\end{lstlisting}

\noindent Location awareness is realized via the special \texttt{context} path element. It
allows to follow the context mapping from platform graph to application graph(s)
and vice versa:

\begin{lstlisting}[language=bash, caption={Location awareness via \texttt{context}.}]
 GET /thread/self/context # on application
 GET /node/42/context     # on platform
\end{lstlisting}

\noindent Accessing \textit{context} from a thread for example will return a processor
core object, accessing it from a core will return a list of thread objects and
so on. Combined with the use of \texttt{parent}, \texttt{self} and
\texttt{context} allows for comprehensive context awareness and exploration.

\subsection{Derived Metrics}

Derived metrics are a core concept of the API as they allow to define high-level
metrics relevant to a specific use-case, user group, experiment, etc. Derived
metrics are generally applied to the  telemetry data on the framework
side, i.e., within the \framework{} service. This allows developers to push
complexity out of their applications. For example, an I/O-sensitive application
might want to terminate or reconfigure if the overall I/O throughput is below a
certain threshold. Instead of periodically querying the I/O metrics for all
processes comprising an application, it is possible to register a derived metric
``I/O Threshold''.

\begin{lstlisting}[language=json, caption={Adding a derived metric on job-level.}]
 PUT /dmetrics
   data {
      metric_name: "i_o_threshold",
      scope: "job",
      function: "..."
   }
\end{lstlisting}

Once a metric is registered, it is available via the \texttt{metrics} section of
the resource object(s) defined in \texttt{scope}. Currently the API does not
come with its own language to define the custom metric \texttt{function}. It
simply uses the query language of the backend system. For our implementation
blueprint explained in more detail in the next section, it uses the functional
expression language used by the Prometheus time series database.

\subsection{Notifications}

Together with derived metrics, notifications are another key concept to address
the endemic pull-based data gathering process found in many applications. The
notification API allows the caller to subscribe to one or more metrics via a
callback mechanism. Whenever the metric changes (beyond a defined threshold),
the callback is engaged. Notifications are user-defined HTTP callbacks,
so-called webhooks. When a new notification is available, the \framework{} API
server makes an HTTP request to the client URI configured for the webhook.

\begin{lstlisting}[language=json, caption={Adding a derived metric on job-level.}]
 PUT /callbacks
   data {
      callback_uri: "http://host/path...",
      scope: "job",
      metric: "i_o_threshold",
   }
\end{lstlisting}

%
%
\section{Implementation Blueprint}
\label{sec:implementation}

\begin{figure*}
  \includegraphics[width=\textwidth]{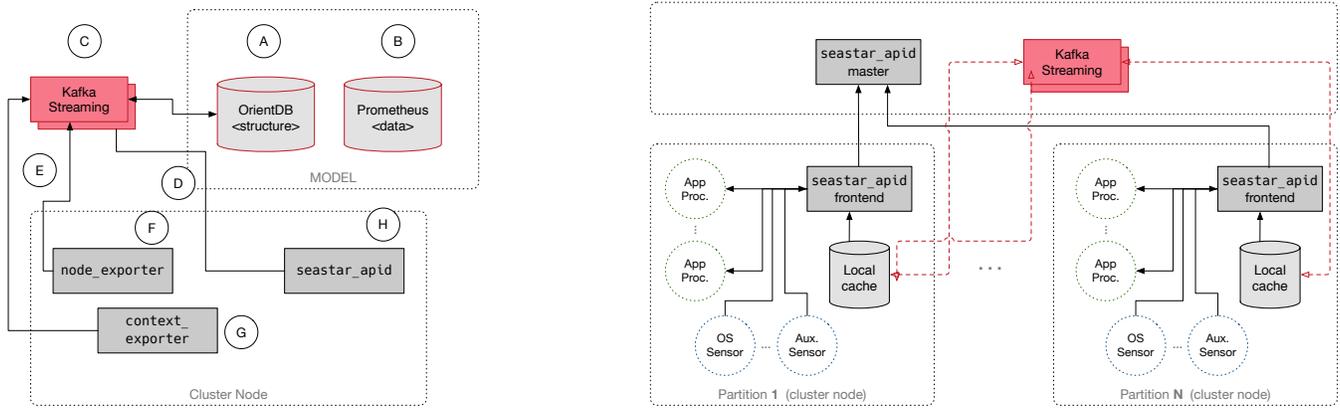}
  \caption{(Left) The \framework{} implementation architecture:
  model databases, data sensors and API services are connected via Kafka. (Right) The API service (\texttt{seastar\_apid}) is
  implemented as a multi-level, partitioned caching architecture to minimize
  telemetry data traffic on the platform. Frontend instances provide the API to
  the consumers via a local cache which is populated with data relevant to the
  instance's partition.}
  \vspace{-1.0em}
  \label{fig:impl_arch}
\end{figure*}

\framework{} tries to be agnostic of applications and platform architectures
and hence does not make many assumptions about how it should be implemented.
In this section, we discuss the \textit{blueprint} for one possible
implementation of \framework{} within an existing HPC cluster. This blueprint
has its origin in the \framework{} research prototype~\cite{seastar2016online}
we have been building to explore various concepts around the API.

In lieu of an actual HPC cluster, our experimental environment
\textit{Elasticluster}~\cite{elasticluster2016online} to start up an on-demand
SLURM-based Linux cluster in the AWS Cloud. This allows us to experiment in
isolation, and also to dynamically change the scale of the cluster. Our
implementation of \framework{} is mostly based on existing technology, not only
to minimize the implementation overhead, but also because there are a plethora
of open-source tools available that provide subsets of the required
functionality at a level of maturity and scalability that would be otherwise
impossible to accomplish.

The implementation architecture (\cref{fig:impl_arch} l.) consists of four main
components: the model server which holds a persistent copy of the
context graph and metrics, the API server which provides the \framework{} API,
and the data sensors, which collect OS, and cluster-level metrics, and the data
backbone which provides a high-throughput, scalable, and buffered data transport
mechanism.

\subsection{Model Database}

The implementation of the \framework{} model is split across two different
databases. A graph-database contains the context, i.e., the spatial-temporal
layout of applications and platform. Another database specialized in storing and
serving large volumes of time-series data efficiently stores the telemetry
data. The node and edge attributes in the graph-database representing the
telemetry data are pointers to the respective entries in the time-series
database. This distinction is not visible in the \framework{} API where
structure and data appear consistent again.

\subsubsection{Context Graph Database}
\label{sec:impl-context-db}

To store the time-variant context graph, we use OrientDB, an open source
multi-model, NoSQL database management system written in Java
(\cref{fig:impl_arch} l. - A). It supports graph, document, key/value, and
object models, with all relationships managed with direct connections between
records.

\subsubsection{Time-Series Database}
\label{sec:impl-time-series-db}

For the time-series database we have chosen Prometheus, an open source
monitoring system and time-series database (\cref{fig:impl_arch} l. - B).
Prometheus can store and process time-series data very efficiently. It  has a
built-in functional expression language that lets the user select  and aggregate
time series data in real time. Furthermore, it has an \textit{Alertmanager}
component which can trigger notifications based on  predefined queries. This
allows for a straight-forward implementation of the derived metrics and
notification functionality of the \framework{} API.

\subsection{Data Transport}
\label{sec:impl-data-sensors}

We use Apache Kafka, an open-source stream processing platform Kafka as the data
transport layer (\cref{fig:impl_arch} l. - C). Kafka provides a
publish-subscribe-based, unified, high-throughput, low-latency platform for
handling real-time data feeds. Kafka makes extensive use of memory channels, and
uses disks as buffers if communication channels are congested or streaming
targets are temporarily not available. This feature adds the necessary
resilience to a distributed system like \framework{}. Kafka can furthermore be
scaled out easily by adding additional nodes. Kafka is responsible for streaming
data in two directions: from the graph- and time-series- databases to the local
API services on the individual cluster nodes (\cref{fig:impl_arch} l. - D) and
from the data sensors to the graph- and time-series- databases
(\cref{fig:impl_arch} l. - E).

\subsection{Data Sensors}
\label{sec:impl-data-sensors}

Data sensors need to capture both, telemetry data as well as the data that is
required to maintain the global context graph, i.e., the relationship between
platform and application. They consist of two components: the
\texttt{node\_exporter} and the \texttt{context\_exporter}. The
\texttt{node\_exporter} (\cref{fig:impl_arch} l. - F) is part of the Prometheus
ecosystem and exports operating-system metrics to the Prometheus server. The
\texttt{context\_exporter} gathers process, job and queueing system information
and sends them to the model database server (\cref{fig:impl_arch} l. - G).

\subsection{API Service}
\label{sec:impl-api}

The API service \texttt{seastar\_apid} (\cref{fig:impl_arch} l. - H) is
implemented as a partitioned caching architecture to minimize network traffic.
(\cref{fig:impl_arch} r.)The service can be instantiated in three different
modes: \textit{master-mode}, \textit{forwarder-mode} and \textit{frontend-mode}.
The frontend instances provide the \framework{} API described in \Cref{sec:api}.
Frontend instances do not have a direct connection to the database, but they
maintain a local data cache which is fed either by an upstream master instance
(2-tier setup) or a forwarder instance (n-tier setup). If a frontend or
forwarder instance cannot serve an API request (cache miss), it sends a request
to its upstream service to provide the missing data set. \texttt{seastar\_apid}
is implemented in Python and uses Python's \textit{FLASK}
HTTP framework. A Python API wrapper provides a more convenient, programmatic
client access to the API service. Especially the well-defined data types free
the user from the burden of parsing JSON return values by hand.

\begin{lstlisting}[language=Python, caption={Python API client}]
from seastar import PlatformAPI

p = PlatformAPI(endpoint='localhost')

rObj = p.self.context.parent
print rObj.kind    # dhcp.type_cpu
print.rObj.metrics # ['memory_total', ... ]

rObj.register_callback(cb_func, ...)
\end{lstlisting}

\noindent The Python API wrapper is only one example of a language-specific wrapper for
the API. Any language for which an HTTP client library exists can interface
with the \framework{} service endpoints. Programming language independence and
the use of standard, well documented protocols fosters adoption of \framework{}
across many different application communities.


\section{Conclusion and Future Work}
\label{sec:conclusion}

In this paper we have picked up the telemetry data management challenge which
we have identified in our previous work~\cite{weidner2016rethinking} as one of
the current challenges in today's HPC ecosystems. We have outlined a solution,
\framework{}, that provides a conceptual framework, and coherent programming
interface for the provisioning and integration of telemetry data on
HPC platforms. We have furthermore sketched out how such a system can be
implemented and integrated with existing HPC platforms. A first prototype
implementation of the model database and API service has the potential to
simplify application development significantly. However, further investigation,
specifically a larger real-world use-case, study still needs to be conducted.

The work presented in this paper is exploratory and the focus has been on
finding the right concepts and abstractions. Future work will focus on the
evaluation of \framework{} and the implementation of application uses
cases.\smallskip{}

\noindent{}\textbf{In-Depth Evaluation:} we will evaluate \framework{} along two
axes: applicability at scale and applicability across different systems. This
will include extensive performance measurements of the suggested architecture
blueprint. The implementation of an adaptive user-level scheduling framework
based on \framework{} as a driving application use-case is already under
development.\smallskip{}

\noindent{}\textbf{Distributed Systems:} many distributed applications strive to
run not just on a single HPC platform but to spread their workload and
components across multiple platforms concurrently. We
will extend the \framework{} model to distributed systems and explore
architectural alternatives for a distributed implementation.\smallskip{}

\noindent{}\textbf{Extreme Scales and Big (Telemetry) Data:} derived metrics
are one of the important concepts in \framework{} to provide telemetry
data to multiple different audiences at different level of abstraction. While
easy enough to manage at small scale, at large scales processing derived metrics
in real time would require a significant amount of computational resources.

\section{Acknowledgments} This research was supported by an \textit{AWS in
Education Research} grant from Amazon Web Services, Inc.

\bibliographystyle{abbrv}
\bibliography{telemetry-data-preprint}

\end{document}